# Chemical principles of instability and self-organization in reacting and diffusive systems

Xiaoliang Wang[1,2]*, Andrew Harrison[3]*

1 College of Life Sciences, Zhejiang University, Hangzhou 310058, China
2 School of Physical Sciences, University of Science and Technology of China, Hefei 230026, China
3 Department of Mathematical Sciences, University of Essex, Colchester CO4 3SQ, UK
*Correspondence: (X.W) wxliang@mail.ustc.edu.cn; (A.H) harry@essex.ac.uk

## Abstract

How patterns and structures undergo symmetry breaking and self-organize within biological systems from initially homogeneous states is a key issue for biological development. The activator-inhibitor (AI) mechanism, derived from reaction-diffusion (RD) models, has been widely believed to be the elementary mechanism for biological pattern formation. This mechanism generally requires activators to be self-enhanced and diffuse more slowly than inhibitors. Here, we identify the instability sources of biological systems and derive the self-organization conditions through solving eigenvalues (dispersion relation) of the generalized RD model for two chemicals. We show that both the single AI mechanisms with long-range inhibition and activation are enough to self-organize into fully-expressed domains without the involvement of the inhibitor-inhibitor (II) mechanism, through singly enhancing the difference in self-proliferation rates of activators and inhibitors or weakening the coupling degree between them. When cross diffusion involves, both the self-enhancement and the difference in diffusion coefficients of chemicals are no longer necessary for self-organization, and the patterning mechanism can be extended to semi-inhibitor and II mechanisms. However, we show that the single activator-activator (AA) mechanism is generally unable to self-organize, even if biological domain growth is additionally involved. Moreover, adding an II system after an AI one can produce discrete and bistable patterns. We also observe that a higher dimensional space can solely alter the patterning principles derived from a lower dimensional space, which may be due to the instability driven by the higher degree of spatial freedom. Such results provide new insights into biological pattern formation.

## 1. Introduction

Self-organized patterning associated with spontaneous symmetry breaking is a hallmark of biological systems at all levels from genomes to ecological systems [1]. In physics, self-organization means a reduction in the uniformity of the system [2]. To understand the underlying universal mechanism has long been of interest to scientists. Turing was the first to propose a reaction-diffusion model to explain biological pattern formation [3]. In the Turing model, the patterning process is reduced to how small spatial fluctuations in a well-mixed system are unstable and become amplified, and a highly ordered and stable periodic structure is finally developed [4]. Mathematical models are necessary in biology [5], since they can reveal the essence behind phenomena.

    Since the publication of Turing's pioneering work in 1952, it has been widely believed that morphogen-mediated "reaction-diffusion" works as the principal mechanism in many embryonic patterning processes [2-21]. In this scenario, complex biological patterns are considered to emerge from the complex interactions among interacting and diffusing chemical molecules (usually termed morphogens), and the genome's role is just to set up the proper initial conditions by its expression of the appropriate set of proteins in the right nuclei.

Thus, in Turing's view, patterns we see in nature are just reflections of the heterogeneities underling biochemical signaling [13]. "Diffusion-driven instability" proposed by Turing has achieved great success in developmental biology and has even been considered as an elementary mechanism for biological patterning.

In our previous work, we have discussed biological pattern formation, using the multiscale discrete element method [22,23]. However, we have not shown the possibility of the continuum theory (the Turing model) in solving key issues of biological development including the precision and robustness of patterning. The Turing model has its unique advantage in theoretical analysis, so we think it should be further stressed.

In this article, we first generalize the formulation of the Turing model (i.e., the deterministic continuum theory) to characterize the interactions between two chemical species, and derive the self-organization conditions. We then discuss the precision of Turing models.

## 2. Generalized Turing model

Under Turing's framework, the biological system we study is abstracted into two reaction-diffusion equations for two chemicals expressed by corresponding genes in cells, which are coupled by the parameter $\lambda$:

$$\begin{aligned}\frac{\partial A}{\partial t} &= D_1 \frac{\partial^2 A}{\partial x^2} + \overbrace{\gamma_1(A - c_0) + \lambda_1(B - c_0)}^{\text{Reaction}} - b(A - c_0)^3 \\ \frac{\partial B}{\partial t} &= D_2 \frac{\partial^2 B}{\partial x^2} + \lambda_2(A - c_0) + \gamma_2(B - c_0) - b(B - c_0)^3\end{aligned} \quad (1)$$

(Diffusion / Reaction / Saturation)

where $D$ is the spatial diffusion constant, $\lambda$ is the coupling rate between A and B, $\gamma$ is the net gain or loss rate, $t$ is time and $x$ is position. $c_0$ is the equilibrium concentration. The last terms in above equations are nonlinear saturation terms, which are added only to constrain chemical concentrations, as done in some literatures.

Obviously, the reaction-diffusion (RD) model is a simplification and omits many details, like the cellular behaviors (e.g., cell growth and division) and the discrete nature of cells, which makes this model only qualitative. Nevertheless, such simplifications are necessary for theorists to obtain analytical solutions and effectively extract the key nature of complex systems.

As in Eq. (1), the self-reproduction rates of chemicals $\gamma_1$ and $\gamma_2$ and the coupling rates between them $\lambda_1$ and $\lambda_2$, constitute a 2×2 topology matrix $\mathcal{T}$ for the interaction network between genes A and B:

$$\mathcal{T} = \begin{pmatrix} \gamma_1 & \lambda_1 \\ \lambda_2 & \gamma_2 \end{pmatrix}$$

The signs of $\lambda_1$ and $\lambda_2$ generally classify the interaction topology into three types, i.e. the activator-inhibitor (AI) mechanism, the inhibitor-inhibitor (II) mechanism (mutual antagonism) and the activator-activator (AA) mechanism (Fig. 1). Here we call the chemical promoting another as the activator, and that inhibiting another as an inhibitor. Different topologies will determine their distinctive properties in pattern formation.

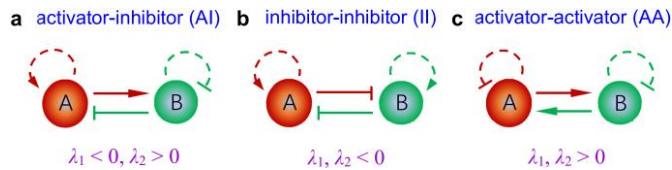

**Figure 1. Topologies of the interaction network between genes A and B.** (**a**) activator-inhibitor (AI) mechanism (A is the activator and B is the inhibitor). (**b**) inhibitor-inhibitor (II) mechanism. (**c**) activator-activator (AA) mechanism. Dash lines represent that they are not essential and even can be replaced by the positive or negative feedback.

### 2.1 Instability of biological system

Biological patterns are initiated by the instability of a homogenous system. The key is to identify instability sources. Eq. (1) is usually analyzed through its linearization [3,24-26]. Given the standing plane-wave solution $c(x,t) = c_0 + a\exp(ikx + wt)$ (Fourier modes, $a$ is the initial amplitude, $k$ is wave number and $w$ is the growth rate of amplitude), the above group of equations can be reduced to an eigenvalue problem:

$$\mathcal{H}\mathbb{C} = w\mathbb{C} \quad (2)$$

where $\mathbb{C} = (a_A, a_B)^T$ and the effective Hamiltonian $\mathcal{H}$ is expressed as follows:

$$\mathcal{H} = \begin{pmatrix} -k^2 D_1 + \gamma_1 & \lambda_1 \\ \lambda_2 & -k^2 D_2 + \gamma_2 \end{pmatrix} \quad (3)$$

We can derive the characteristic equation that corresponds to $\mathcal{H}$ as:

$$m_1 \cdot m_2 - 1 = 0 \quad (4)$$

where $m_1$ and $m_2$ are expressed as:

$$m_1 = \frac{w + k^2 D_1 - \gamma_1}{\lambda_1}, \quad m_2 = \frac{w + k^2 D_2 - \gamma_2}{\lambda_2} \quad (5)$$

Combining Eqs. (4) and (5), the eigenvalue of this system (i.e. the dispersion relation) can be finally solved as:

$$w_{\pm} = -\frac{k^2(D_1+D_2) - (\gamma_1+\gamma_2)}{2} \pm \sqrt{\lambda_1\lambda_2 + \left[\frac{k^2(D_1-D_2) + (\gamma_2-\gamma_1)}{2}\right]^2} \quad (6)$$

Form Eq. (6), we can obtain the maximum growth rate of instability as:

$$\sigma = \frac{(\gamma_1+\gamma_2) - k^2(D_1+D_2)}{2} + \sqrt{\lambda_1\lambda_2 + \left[\frac{k^2(D_1-D_2) + (\gamma_2-\gamma_1)}{2}\right]^2} \quad (7)$$

where $\lambda_1\lambda_2 + \left[\frac{k^2(D_1-D_2) + (\gamma_2-\gamma_1)}{2}\right]^2 > 0$.

Fourier analysis here shows that the instability of the system of two reacting and diffusing chemicals could come from (1) the self-enhancement of chemicals (i.e., the self-proliferation rate $\gamma_{1,2} > 0$), (2) the difference in self-reproduction rates $|\gamma_1 - \gamma_2|$, (3) the difference in diffusion coefficients $|D_1 - D_2|$ and (4) the positive product of coupling rates $\lambda_1\lambda_2 > 0$, which can all serve as the sources of symmetry breaking of a biological system.

According to Eq. (7), the most unstable wave happens at $\partial\sigma(k)/\partial k = 0$, namely

$$D_1 D_2 A^2 + \lambda_1\lambda_2(D_1+D_2)^2 = 0 \quad (8)$$

where $A = k^2(D_1 - D_2) + (\gamma_2 - \gamma_1)$.

Eq. (8) has non-zero/nontrivial solutions only when $\lambda_1\lambda_2 < 0$, namely for AI systems. In this case, the solutions $k_m$ satisfies

$$k_m^2 = \frac{(\gamma_1 - \gamma_2) \pm (D_1+D_2)\sqrt{-\frac{\lambda_1\lambda_2}{D_1 D_2}}}{D_1 - D_2} \quad (9)$$

The final amplitude of standing wave can be accordingly determined as $a_m = \sqrt{\frac{\sigma_m}{b}}$, where $\sigma_m$ is the growth rate of instability of the most unstable mode.

For AA and II systems ($\lambda_1\lambda_2 > 0$), $\sigma(k)$ is generally the monotonic function of wave number $k$, which leads to no stable waves within the system (the wave number $k_m$ is zero).

To stabilize the most unstable mode, it is required that $\sigma(k_m) > max(0, Re(\sigma(0)))$ and $k_m^2 > 0$. According to Eqs. (7) and (9), we have,

$$\begin{cases} \sigma(k_m) = \dfrac{(\gamma_1+\gamma_2) - k_m^2(D_1+D_2)}{2} + \dfrac{|D_1 - D_2|}{2}\sqrt{-\dfrac{\lambda_1\lambda_2}{D_1D_2}} > 0 \\ \sigma(k_m) - Re(\sigma(0)) = \dfrac{-k_m^2(D_1+D_2)}{2} + \dfrac{|D_1 - D_2|}{2}\sqrt{-\dfrac{\lambda_1\lambda_2}{D_1D_2}} - Re\left(\sqrt{\lambda_1\lambda_2+\dfrac{(\gamma_1 - \gamma_2)^2}{4}}\right) > 0 \\ k_m^2 > 0 \end{cases} \quad (10)$$

From the inequality (10), we further have,

$$\begin{cases} \dfrac{(\gamma_1 + \gamma_2) + |D_1 - D_2|\sqrt{-\dfrac{\lambda_1\lambda_2}{D_1D_2}}}{D_1+D_2} > k_m^2 > 0 \\ \dfrac{|D_1 - D_2|\sqrt{-\dfrac{\lambda_1\lambda_2}{D_1D_2}} - Re\left(\sqrt{4\lambda_1\lambda_2+(\gamma_1 - \gamma_2)^2}\right)}{D_1+D_2} > k_m^2 > 0 \end{cases} \quad (11)$$

Combining Eq. (9) with the inequality (11), we can derive the self-organization condition (the parameter range):

$$-\dfrac{(D_1 - D_2)(\gamma_1 - \gamma_2)D_1D_2}{D_1+D_2} < |D_1 - D_2|\sqrt{-\lambda_1\lambda_2 D_1D_2} < min\left(\dfrac{(D_1\gamma_2 - D_2\gamma_1)(D_1 - D_2)}{2}, -\dfrac{(D_1^2 - D_2^2)(\gamma_1 - \gamma_2)+(D_1 - D_2)^2 Re\left(\sqrt{4\lambda_1\lambda_2+(\gamma_1 - \gamma_2)^2}\right)}{4}\right) \quad (12)$$

The inequility (12) suggests that the difference in diffusion coefficients of chemicals is necessary for patterning. Besides, the exchage of the signs of $\lambda_1$ and $\lambda_2$ has no impact on stabilization conditions, no matter $D_1 < D_2$ or $D_1 > D_2$. This indicates in theory that if the long-range ihibition is able to self-organize, then the long-range activation is also capable. For example, the case $D_1 < D_2$ of the inequility (12) suggests that,

$$\begin{cases} \dfrac{D_2\gamma_1 - D_1\gamma_2}{2} > 0 \\ \dfrac{(D_1+D_2)(\gamma_1 - \gamma_2)}{4} > 0 \\ \dfrac{D_2\gamma_1 - D_1\gamma_2}{2} > \dfrac{(\gamma_1 - \gamma_2)D_1D_2}{D_1+D_2} \end{cases} \quad (13)$$

From the inequility (13), we can derive that $\gamma_1 > max\left(\gamma_2, -\dfrac{D_1}{D_2}\gamma_2\right) \geq 0$ is required for self-organizing into Turing patterns (i.e. the standing wave with a single mode). This means that for the long-range ihibition mechanism ($\lambda_1 < 0$ and $\lambda_2 > 0$, the classical Turing mechanism), the activator is required to be self-enhanced, and for the long-range activation mechanism ($\lambda_1 > 0$ and $\lambda_2 < 0$), the inhibitor is required to be self-enhanced. This conclusion can also be obtained through analyzing the case $D_1 > D_2$ of the inequility (12).

Fig. 2 shows the self-organization of the AI mechanisms with long-range inhibition and long-range activation within one-dimensional stationary domains. Under the long-range inhibition, chemicals A and B are in identical phase, while chemicals A and B are in opposed phase under the long-range activation which is also termed the "activator-substrate" (AS) mechanism [25,27]. Especially, when the instability strength of unstable waves $\sigma(k)$ is enhanced due to higher $\gamma_1$ and $\gamma_2$ and the weaker coupling (larger $\lambda_1\lambda_2$) of chemicals (see Eq. 7), there is also a possibility for such an AI mechanism to give rise to flat patterns of the activator A or the inhibitor B which tend to only occur with the involvement of the II mechanism [28]. It seems to demonstrate that some special pattern formation functions that are similar to that of the II system may be achieved by tuning singly the AI system. This exception is missed in a recent classification discussion of the 2-node network (type I) [28]. If $\sigma(k)$ is in high enough level, the AI mechanism is further observed to give rise to less regular patterns, although the most unstable mode in theory is $k_m > 0$. The irregularity of patterns may mean that high $\sigma(k)$ has caused the most unstable mode hard to fully suppress the other modes before chemicals at some positions reach their saturation points, leading multiple unstable wave modes to finally stabilize within the system.

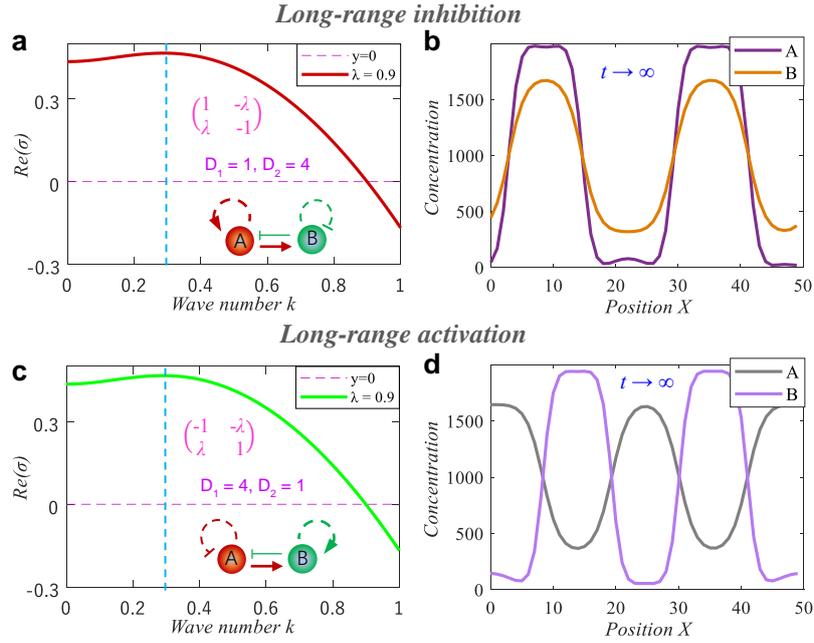

**Figure 2. Pattern formation of the single AI mechanism on stationary 1D domains.** (**a**), (**b**) The AI mechanism with long-range inhibition ($D_1 = 1$, $D_2 = 4$): (**a**), Growth rate of the unstable waves as the function of the wave number; (**b**), The final unstable waves of chemicals A and B within the system ($b = 4.1 \times 10^{-7}$): A and B are in identical phase. (**c**), (**d**) The AI mechanism with long-range activation ($D_1 = 4$, $D_2 = 1$): (**c**), Growth rate of the unstable waves as the function of the wave number; (**d**), The final unstable waves of chemicals A and B within the system ($b = 4.5 \times 10^{-7}$) : A and B are in opposed phase. Periodic boundary conditions are used.

Our results also suggest that tuning solely the self-proliferation rates, the coupling degree and diffusion speed of interacting chemicals can realize the scaling of Turing patterns, just as presented by Eq. (9).

**Table 1: The AI mechanism in organisms identified by molecular experiments.**

| Genetic interaction mode | Identified interactions | Functions |
|---|---|---|
| AI mechanism | Wnt ⊣ Dkk | a. Lung branching in mouse[29]<br>b. Head regeneration in Hydra[30]<br>c. Hair follicle spacing in mouse[31] |
| | Nodal ⊣ Lefty | a. Antero-posterior patterning in zebrafish[32]<br>b. Left-right asymmetry in limbs[33] |
| | FGF ⊣ BMP | Hair regeneration in mouse[34] |

As summarized in Table 1, the activator-inhibitor mechanism with long-range inhibition has been claimed to be identified in various embryonic development processes, such as Wnt-Dkk [29-31], Nodal-Lefty [32,33] and FGF-BMP [34] pairs, including the head regeneration in Hydra, hair follicle spacing in mouse, and the left-right asymmetry in mouse embryos. Nevertheless, this mechanism has not been fully recognized by experimental researchers [35].

## 2.2 Effects of domain growth and cross diffusion

Both the domain growth of biological systems and the cross diffusion of chemicals have been observed to play an important role in biological pattern formation [36-42]. Here we further stress these two issues. Taking domain growth and cross diffusion into generalized Eq. (1), we have:

$$\begin{aligned}\frac{\partial A}{\partial t} + h(t)A &= \frac{1}{\varphi^2(t)}\left(D_1\frac{\partial^2 A}{\partial x^2} + D_{12}\frac{\partial^2 B}{\partial x^2}\right) + \overbrace{\gamma_1(A - c_0) + \lambda_1(B - c_0)}^{\text{Reaction}} - b(A - c_0)^3 \\ \frac{\partial B}{\partial t} + h(t)B &= \frac{1}{\varphi^2(t)}\left(D_2\frac{\partial^2 B}{\partial x^2} + D_{21}\frac{\partial^2 A}{\partial x^2}\right) + \lambda_2(A - c_0) + \gamma_2(B - c_0) - b(B - c_0)^3\end{aligned} \quad (14)$$

where $h(t)$ and $\varphi(t)$ respectively represent the dilution effect and spatial expansion effect induced by domain growth. For the linear and exponential domain growth, $h(t)$ and $\varphi(t)$ are listed in Table 2.

Table 2: Functions $h(t)$ and $\varphi(t)$ for the linear and exponential domain growth.

| Growth type | $\varphi(t)$ | $h(t)$ |
| --- | --- | --- |
| Linear | $\varphi(t) = 1 + \frac{vt}{L_0}$ | $h(t) = \frac{mv}{L_0 + vt}$ |
| Exponential | $\varphi(t) = e^{vt}$ | $h(t) = mv$ |

$L_0$ is the initial domain length, and $m$ is the number of spatial dimensions.

The effective Hamiltonian $\mathcal{H}$ for this system is:

$$\mathcal{H} = \begin{pmatrix} -\frac{k^2 D_1}{\varphi^2(t)} + \gamma_1 - h(t) & -\frac{k^2 D_{12}}{\varphi^2(t)} + \lambda_1 \\ -\frac{k^2 D_{21}}{\varphi^2(t)} + \lambda_2 & -\frac{k^2 D_2}{\varphi^2(t)} + \gamma_2 - h(t) \end{pmatrix} \quad (15)$$

Accordingly, we can derive the maximum growth rate of instability:

$$\sigma = \frac{(\gamma_1 + \gamma_2) - \frac{k^2}{\varphi^2(t)}(D_1 + D_2)}{2} - h(t) + \sqrt{\left(\lambda_1 - \frac{k^2 D_{12}}{\varphi^2(t)}\right)\left(\lambda_2 - \frac{k^2 D_{21}}{\varphi^2(t)}\right) + \left[\frac{\frac{k^2}{\varphi^2(t)}(D_1 - D_2) + (\gamma_2 - \gamma_1)}{2}\right]^2} \quad (16)$$

❒ For simplicity, we now first consider the absence of domain growth, $D_1 = D_2 = D$, $D_{12} = D_{21} = D'$ and $\gamma_1 = \gamma_2 = \gamma$, then Eq. (16) becomes

$$\sigma = \gamma - k^2 D + \sqrt{\left[D'k^2 - \frac{(\lambda_1 + \lambda_2)}{2}\right]^2 - \left(\frac{\lambda_1 - \lambda_2}{2}\right)^2} \quad (17)$$

According to Eq. (17), we have $\partial\sigma(k)/\partial k = 0$, namely,

$$\left(D^2 - D'^2\right)A^2 - \frac{D^2(\lambda_1 - \lambda_2)^2}{4} = 0 \qquad (18)$$

where $A = D'k^2 - \frac{(\lambda_1+\lambda_2)}{2}$.

Eq. (18) has non-zero solutions only when $D > D'$, and the solutions satisfy

$$k_m^2 = \frac{(\lambda_1+\lambda_2) \pm D(\lambda_1 - \lambda_2)/\sqrt{D^2 - D'^2}}{2D'} \qquad (19)$$

Accordingly, we can derive the self-organization conditions for such a case as follows:

$$-\frac{(\lambda_1+\lambda_2)\left(D^2 - D'^2\right)}{D} < |\lambda_1 - \lambda_2|\sqrt{D^2 - D'^2} < \min\left(2D'\gamma - D(\lambda_1+\lambda_2), -D(\lambda_1+\lambda_2) - 2D'Re(\sqrt{\lambda_1\lambda_2})\right) \qquad (20)$$

Compared with the inequality (12), the inequality (20) suggests that both the activator-inhibitor mechanism and self-enhancement of chemicals ($\gamma_{1,2} > 0$) are not required for self-organization. The inequality (20) also suggests that,

$$\begin{cases} 2D'\gamma - D(\lambda_1+\lambda_2) > 0 \\ -D(\lambda_1+\lambda_2) > 0 \\ 2D'\gamma - D(\lambda_1+\lambda_2) > -\frac{(\lambda_1+\lambda_2)\left(D^2 - D'^2\right)}{D} \end{cases} \qquad (21)$$

From the inequility (21), we can derive that $\lambda_1+\lambda_2 < \min\left(0, 2\frac{D'}{D}\gamma\right) \leq 0$ is required for self-organization. The requirement of $\lambda_1+\lambda_2 < 0$ suggests that for the current case, the activator-inhibitor, inhibitor-inhibitor and semi-inhibitor (i.e. $\lambda_1 = 0, \lambda_2 < 0$ or $\lambda_2 = 0, \lambda_1 < 0$) mechanisms are all potential to self-organize, while the single activator-activator mechanism is infeasible ($\lambda_1+\lambda_2 > 0$).

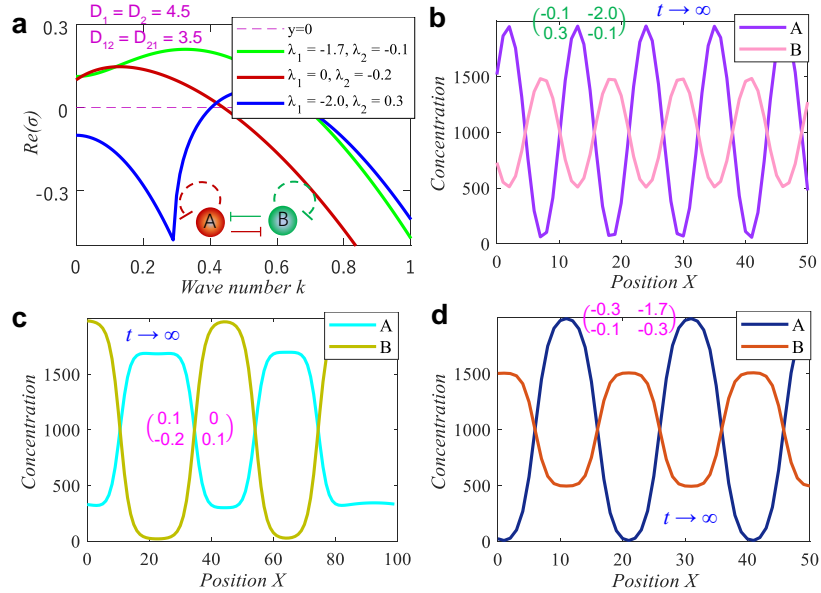

**Figure 3. Pattern formation of the single AI, semi-inhibitor and II mechanisms in the presence of cross diffusion.** (**a**), Growth rate of the unstable waves as the function of the wave number; (**b**), (**c**), (**d**) The final unstable waves of chemicals A and B within the system ($b$ = 1.4×10$^{-7}$ for **b**, $b$ = 2.4×10$^{-7}$ for **c** and $b$ = 4.2×10$^{-7}$ for **d**). D$_1$=D$_2$=4.5 and D$_{12}$ = D$_{21}$ = 3.5. Periodic

boundary conditions are used.

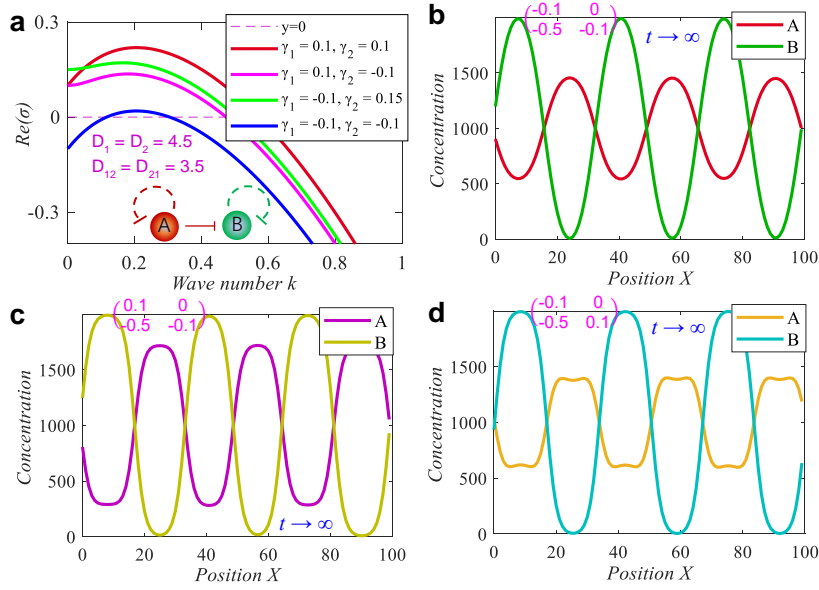

**Figure 4. Pattern formation of the semi-inhibitor mechanism with different topologies in the presence of cross diffusion.** (**a**), Growth rate of the unstable waves as the function of the wave number; (**b**), (**c**), (**d**) The final unstable waves of chemicals A and B within the system ($b = 0.4\times10^{-7}$ for **b**, $b = 2.4\times10^{-7}$ for **c** and **d**). $D_1=D_2=4.5$ and $D_{12} = D_{21} = 3.5$. Periodic boundary conditions are used.

Fig. 3 shows the self-organization of single AI, semi-inhibitor and II mechanisms in the presence of cross diffusion. Results demonstrate that when the cross diffusion of chemicals is involved, both the difference in diffusion and the self-enhancement of chemicals are no longer necessary for self-organization of the single AI mechanism, and the self-organization mechanism can be extended to II and semi-inhibitor mechanisms. Fig. 4 shows that various topologies of the semi-inhibitor mechanism, i.e. $\gamma_1\gamma_2 > 0$, $\gamma_1\gamma_2 < 0$ and even $\gamma_1\gamma_2 = 0$ are all able to patterning. This result is also observed to hold for the AI and II mechanisms.

☐ For the more general case $D_{12} \neq D_{21}$, we have

$$\sigma = \gamma - k^2 D + \sqrt{\left[\sqrt{D_{12}D_{21}}k^2 - \frac{(\lambda_1 D_{21}+\lambda_2 D_{12})}{2\sqrt{D_{12}D_{21}}}\right]^2 - \frac{(\lambda_1 D_{21} - \lambda_2 D_{12})^2}{4D_{12}D_{21}}} \quad (22)$$

The most unstable wave mode occurs at $\partial\sigma(k)/\partial k = 0$, namely,

$$(D^2 - D_{12}D_{21})A^2 - \frac{D^2(\lambda_1 D_{21} - \lambda_2 D_{12})^2}{4D_{12}D_{21}} = 0 \quad (23)$$

where $A = \sqrt{D_{12}D_{21}}k^2 - \frac{(\lambda_1 D_{21}+\lambda_2 D_{12})}{2\sqrt{D_{12}D_{21}}}$.

Eq. (23) has non-zero solutions only when $D^2 > D_{12}D_{21}$, and the solutions satisfy

$$k_m^2 = \frac{(\lambda_1 D_{21}+\lambda_2 D_{12}) \pm D(\lambda_1 D_{21} - \lambda_2 D_{12})/\sqrt{D^2 - D_{12}D_{21}}}{2D_{12}D_{21}} \quad (24)$$

Accordingly, we can derive the self-organization conditions for such a case as follows:

$$-\frac{(\lambda_1 D_{21}+\lambda_2 D_{12})(D^2 - D_{12}D_{21})}{D} < |\lambda_1 D_{21} - \lambda_2 D_{12}|\sqrt{D^2 - D_{12}D_{21}} < \min\left(2D_{12}D_{21}\gamma - D(\lambda_1 D_{21}+\lambda_2 D_{12}), -D(\lambda_1 D_{21}+\lambda_2 D_{12}) - 2D_{12}D_{21}Re(\sqrt{\lambda_1\lambda_2})\right) \quad (25)$$

For the more general case $D_{12} \neq D_{21}$, we can derive in the same way that $\lambda_1 D_{21}+\lambda_2 D_{12} < 0$ (exactly $\lambda_1 D_{21}+\lambda_2 D_{12} < min[0, 2D\gamma] \leq 0$) is necessary for self-organization, which also means the AA mechanism is infeasible except the AI and II mechanisms.

❏ For the more general case $D_1 \neq D_2$ and $D_{12} \neq D_{21}$, we have

$$\sigma = \gamma - \frac{1}{2}(D_1+D_2)k^2 + \sqrt{\left[\sqrt{D_{12}D_{21}+\frac{1}{4}(D_1-D_2)^2}k^2 - \frac{(\lambda_1 D_{21}+\lambda_2 D_{12})}{\sqrt{4D_{12}D_{21}+(D_1-D_2)^2}}\right]^2 - \frac{(\lambda_1 D_{21}-\lambda_2 D_{12})^2 - \lambda_1\lambda_2(D_1-D_2)^2}{4D_{12}D_{21}+(D_1-D_2)^2}} \quad (26)$$

The most unstable wave mode occurs at $\partial\sigma(k)/\partial k = 0$, namely,

$$4(D_1 D_2 - D_{12}D_{21})A^2 - \frac{(D_1+D_2)^2[(\lambda_1 D_{21}-\lambda_2 D_{12})^2 - \lambda_1\lambda_2(D_1-D_2)^2]}{4D_{12}D_{21}+(D_1-D_2)^2} = 0 \quad (27)$$

where $A = \sqrt{D_{12}D_{21}+\frac{1}{4}(D_1-D_2)^2}k^2 - \frac{(\lambda_1 D_{21}+\lambda_2 D_{12})}{\sqrt{4D_{12}D_{21}+(D_1-D_2)^2}}$.

Eq. (27) has non-zero solutions only when $\frac{(\lambda_1 D_{21}-\lambda_2 D_{12})^2 - \lambda_1\lambda_2(D_1-D_2)^2}{D_1 D_2 - D_{12}D_{21}} > 0$, and the solutions satisfy

$$k_m^2 = \frac{2(\lambda_1 D_{21}+\lambda_2 D_{12}) \pm (D_1+D_2)\sqrt{\frac{(\lambda_1 D_{21}-\lambda_2 D_{12})^2 - \lambda_1\lambda_2(D_1-D_2)^2}{D_1 D_2 - D_{12}D_{21}}}}{4D_{12}D_{21}+(D_1-D_2)^2} \quad (28)$$

Accordingly, self-organization conditions can be obtained as:

$$-\frac{4(\lambda_1 D_{21}+\lambda_2 D_{12})(D_1 D_2 - D_{12}D_{21})}{D_1+D_2} < 2\sqrt{(D_1 D_2 - D_{12}D_{21})[(\lambda_1 D_{21}-\lambda_2 D_{12})^2 - \lambda_1\lambda_2(D_1-D_2)^2]} < min\left([4D_{12}D_{21}+(D_1-D_2)^2]\gamma - \right.$$

$$\left. (D_1+D_2)(\lambda_1 D_{21}+\lambda_2 D_{12}), -(D_1+D_2)(\lambda_1 D_{21}+\lambda_2 D_{12}) - [4D_{12}D_{21}+(D_1-D_2)^2]Re(\sqrt{\lambda_1\lambda_2})\right) \quad (29)$$

For the more general case $D_1 \neq D_2$ and $D_{12} \neq D_{21}$, it is still derived that $\lambda_1 D_{21}+\lambda_2 D_{12} < min[0, (D_1+D_2)\gamma] \leq 0$ ($\lambda_1 D_{21}+\lambda_2 D_{12} < 0$) is necessary, and the AA mechanism is unable to self-organize.

❏ When we consider the case $D_1 \neq D_2$ $D_{12} \neq D_{21}$ and $\gamma_1 \neq \gamma_2$, we have,

$$\sigma = \frac{(\gamma_1+\gamma_2) - k^2(D_1+D_2)}{2} +$$

$$\sqrt{\left[\sqrt{D_{12}D_{21}+\frac{1}{4}(D_1-D_2)^2}k^2 - \frac{(\lambda_1 D_{21}+\lambda_2 D_{12})+\frac{1}{2}(D_1-D_2)(\gamma_1-\gamma_2)}{\sqrt{4D_{12}D_{21}+(D_1-D_2)^2}}\right]^2 + \lambda_1\lambda_2 + \frac{(\gamma_1-\gamma_2)^2}{4} - \frac{\left[(\lambda_1 D_{21}+\lambda_2 D_{12})+\frac{1}{2}(D_1-D_2)(\gamma_1-\gamma_2)\right]^2}{4D_{12}D_{21}+(D_1-D_2)^2}} \quad (30)$$

The most unstable wave mode occurs at $\partial\sigma(k)/\partial k = 0$, i.e.

$$4(D_1 D_2 - D_{12}D_{21})A^2 + (D_1+D_2)^2\left[\lambda_1\lambda_2 + \frac{(\gamma_1-\gamma_2)^2}{4} - \frac{\left[(\lambda_1 D_{21}+\lambda_2 D_{12})+\frac{1}{2}(D_1-D_2)(\gamma_1-\gamma_2)\right]^2}{4D_{12}D_{21}+(D_1-D_2)^2}\right] = 0 \quad (31)$$

where $A = \sqrt{D_{12}D_{21}+\frac{1}{4}(D_1-D_2)^2}k^2 - \frac{(\lambda_1 D_{21}+\lambda_2 D_{12})+\frac{1}{2}(D_1-D_2)(\gamma_1-\gamma_2)}{\sqrt{4D_{12}D_{21}+(D_1-D_2)^2}}$.

Eq. (31) have non-zero solutions only when $B^2 - [4D_{12}D_{21}+(D_1-D_2)^2]\left[\lambda_1\lambda_2 + \frac{1}{4}(\gamma_1-\gamma_2)^2\right] > 0$, and we will obtain

$$k_m^2 = \frac{2B \pm (D_1+D_2)\sqrt{\frac{B^2 - [4D_{12}D_{21}+(D_1-D_2)^2]\left[\lambda_1\lambda_2 + \frac{1}{4}(\gamma_1-\gamma_2)^2\right]}{D_1 D_2 - D_{12}D_{21}}}}{4D_{12}D_{21}+(D_1-D_2)^2} \quad (32)$$

where $B = (\lambda_1 D_{21} + \lambda_2 D_{12}) + \frac{1}{2}(D_1 - D_2)(\gamma_1 - \gamma_2)$.

Accordingly, self-organization conditions can be obtained as:

$$-\frac{4B(D_1 D_2 - D_{12} D_{21})}{D_1 + D_2} < 2\sqrt{(D_1 D_2 - D_{12} D_{21})\left[B^2 - [4D_{12} D_{21} + (D_1 - D_2)^2]\left[\lambda_1 \lambda_2 + \frac{1}{4}(\gamma_1 - \gamma_2)^2\right]\right]} < \min\left([4D_{12} D_{21} + (D_1 - D_2)^2]\frac{(\gamma_1 + \gamma_2)}{2} - (D_1 + D_2)B, -(D_1 + D_2)B - [4D_{12} D_{21} + (D_1 - D_2)^2]Re\left(\sqrt{\lambda_1 \lambda_2 + \frac{1}{4}(\gamma_1 - \gamma_2)^2}\right)\right) \quad (33)$$

We will derive in the same way that $\lambda_1 D_{21} + \lambda_2 D_{12} + \frac{1}{2}(D_1 - D_2)(\gamma_1 - \gamma_2) < 0$ ($B < \min\left[0, (D_1 + D_2)\frac{(\gamma_1 + \gamma_2)}{2}\right] \leq 0$) is necessary for self-organization.

To confirm whether the AA mechanism is able to patterning for the current case, we first assume there exists one AA case ($\lambda_{1,2} > 0$) that can self-organize. From $B^2 - [4D_{12} D_{21} + (D_1 - D_2)^2]\left[\lambda_1 \lambda_2 + \frac{1}{4}(\gamma_1 - \gamma_2)^2\right] > 0$, we can further obtain

$$(\lambda_1 D_{21} + \lambda_2 D_{12})(D_1 - D_2)(\gamma_1 - \gamma_2) > -(\lambda_1 D_{21} - \lambda_2 D_{12})^2 + \lambda_1 \lambda_2 (D_1 - D_2)^2 + D_1 D_2 (\gamma_1 - \gamma_2)^2 \quad (34)$$

When $\lambda_{1,2} > 0$, $\lambda_1 D_{21} + \lambda_2 D_{12} > 0$. So we can derive from $B < 0$ that
$$(\lambda_1 D_{21} + \lambda_2 D_{12})(D_1 - D_2)(\gamma_1 - \gamma_2) < -2(\lambda_1 D_{21} + \lambda_2 D_{12})^2 \quad (35)$$

Inequalities (34) and (35) suggest it is required that

$$-2(\lambda_1 D_{21} + \lambda_2 D_{12})^2 > -(\lambda_1 D_{21} - \lambda_2 D_{12})^2 + \lambda_1 \lambda_2 (D_1 - D_2)^2 + D_1 D_2 (\gamma_1 - \gamma_2)^2$$

Namely, it is required that

$$(\lambda_1 D_{21} + \lambda_2 D_{12})^2 + \lambda_1 \lambda_2 [4D_{12} D_{21} + (D_1 - D_2)^2] + D_1 D_2 (\gamma_1 - \gamma_2)^2 < 0 \quad (36)$$

However, when $\lambda_{1,2} > 0$, $\lambda_1 \lambda_2 [4D_{12} D_{21} + (D_1 - D_2)^2] > 0$, which means the inequality (36) is invalid. Therefore, for the current case, the AA mechanism is also impossible to self-organize.

In conclusion, our mathematical analysis generally suggests that in the presence of cross diffusion, the self-organization mechanism of a stationary RD system can only be extended to semi-inhibitor ($\lambda_1 = 0, \lambda_2 < 0$ or $\lambda_2 = 0, \lambda_1 < 0$) and inhibitor-inhibitor ($\lambda_{1,2} < 0$) mechanisms, while the activator-activator ($\lambda_{1,2} > 0$) mechanism is still impossible to self-organize. We can also further deduce that these conclusions will hold even if the domain growth is additionally involved.

❒ We now consider the effect of domain growth.

According to Eq. (32), it can be derived that $k_m \propto \varphi(t)$ when domain growth is involved. Since $\varphi(t) = L(t)/L_0$ ($L(t)$ is the domain length at time $t$), we will get

$$\lambda_m L(t) = const. \quad (37)$$

where $\lambda_m = 2\pi/k_m$ is the wave length of the most unstable mode.

Eq. (36) suggests that for one given growing biological system, the periodic length of produced ordered structures generally declines with time and is in reverse proportion to the size of the system. This phenomenon has been presented in theoretical work [37,38].

## 3. Precise and robust pattern formation on 2D domains

The precise positioning of gene expression domains in cell populations is a big challenge for the current understanding of developmental pattern formation [9,43]. As presented above, patterns produced by the RD mechanism are generally continuously graded and have no well-defined region boundaries. In morphogenesis, however, cells' fates are often controlled very precisely at specific positions. Take the zebra stripe shown in Fig. 5a for an example. There are domains composed of pigment cells which are completely segregated in space, and one cannot find a different pigment within another pigment's domain. The RD mechanism has been questioned because of this problem, since the graded morphogen gradients generated by diffusion are believed to be too messy to drive the tissue patterning in organisms [9].

To explain the formation of patterns with well-defined (discontinuous) domains, the French flag model [44] has been proposed. In this model, it is postulated that different pigment genes are activated by a graded morphogen distribution providing different position information in a concentration-dependent manner with corresponding thresholds [9] (Fig. 5b). Cells will adopt different fates at the signaling levels which are below and above the thresholds [8]. Although many morphogens with several thresholds of gene expression induction have been identified by experiments [45-49], it is still a puzzle for this model why one gene cannot be activated on the higher morphogen (transcriptional factor) concentration if it can be activated at a lower level of concentration. It is also hard to explain the generation of periodic patterns, as in Fig. 5a, with the French flag model.

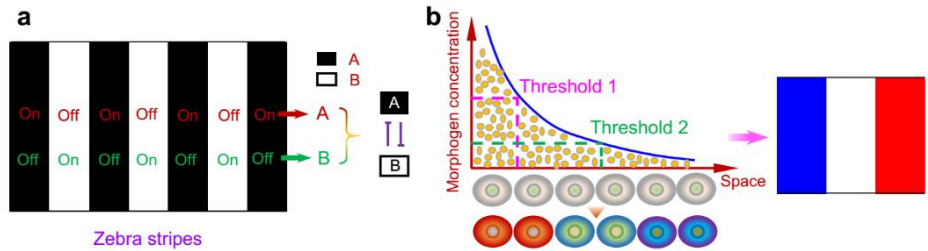

**Figure 5.** (**a**) Illustration of zebra stripes with discrete domains: The mutual antagonism between expression states of white and black genes probably indicates the involvement of II mechanism in pattern formation. (**b**) Interpretation of morphogen signals in French flag model: Morphogen gradient provides the position information to cells and makes them transform the information into the final region-specific fates.

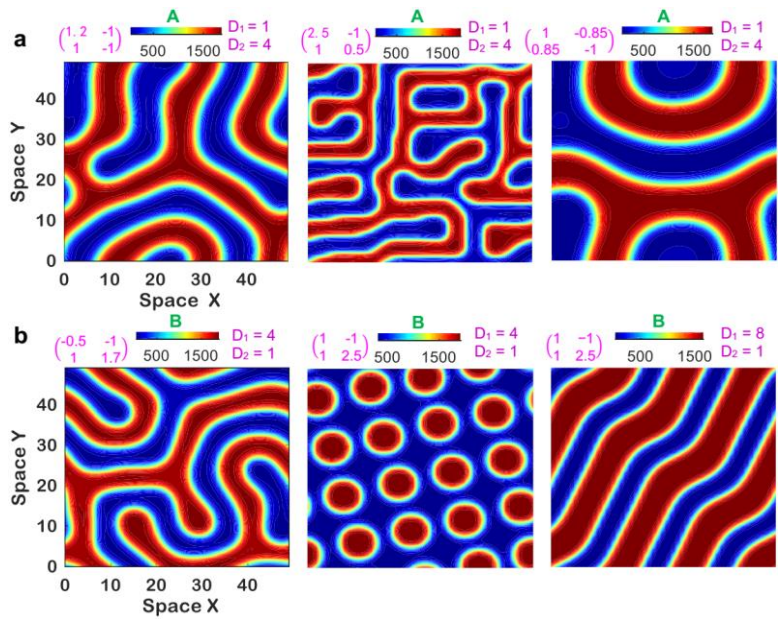

**Figure 6. Precise pattern formation of the single AI mechanism on stationary 2D domains.** (**a**), The AI mechanism with long-range inhibition: spatial distribution of the chemical A ($b = 6.8 \times 10^{-7}$ for the left, $b = 20.9 \times 10^{-7}$ for the middle and $b = 5.7 \times$

$10^{-7}$ for the right); (**b**) The AI mechanism with long-range activation: spatial distribution of the chemical B ($b$ =16×$10^{-7}$ for the left, $b$ = 27.5×$10^{-7}$ for the middle and $b$ = 28.5×$10^{-7}$ for the right). Patterns are produced with the 2D form of Eq. (1). Periodic boundary conditions are used.

We believe that the precision problem of the Turing model does not root in diffusion itself but in the single selection of the AI mechanism with inappropriate topological parameters. Fig. 6 presents pattern formation on two dimensional (2D) stationary domains, using the same Eq. (1). It further shows that the single AI mechanism can obtain precise stable stripe and spot patterns with well-defined domains through tuning topology parameters, including elevating self-proliferation rates of chemicals and weakening the coupling degree between activators and inhibitors. Within a 2D space, a RD system usually has a high sensitivity to the details of both initial and boundary conditions, and can produce a nearly limitless variety of spatial patterns, through singly tuning model parameters and boundary conditions [13,17,35,50]. The remarkable similarity of some patterns with those seen in real organisms has been presented by Kondo and Miura [35].

The antagonistic relationship between white and black stripes in the zebra stripes shown in Fig. 5a probably indicates the involvement of the II mechanism, which may take a part in precise pattern formation. We verified this deduction through adding an II system after an AI one (Fig. 7). We use the patterned structures resulting from the AI system as the signal inputs of the II system. It is observed that the sequential combination between AI and II mechanisms has dramatically enhanced the precision of patterning. At the same time, chemicals C and D suppress each other in full within their own dominant expression domains, as disclosed in our previous study [23].

Within the designed sequential combination of AI and II mechanisms in Fig. 7a, the II subsystem is started after the AI subsystem with a time delay Δt. Fig. 7d shows that if the time delay is too short, there will be an obvious error for the gene C to interpret signals from the gene A at some positions, since the II system starts before the AI system has not reached the steady state. Our results also demonstrate that longer time delay is required for precise patterning if the mutual inhibition of the II subsystem is stronger (i.e. higher $\lambda_3\lambda_4$). Figs. 7e and 7f suggest that the patterning precision can be improved through elevating the mutual inhibition degree of the II system and the activation degree from the AI system. Specifically, Fig. 7e shows that patterning can still be refined to some extent even if there exists no mutual inhibition between activated genes C and D. It is shown in Fig. 7f that activation signals A and B have significant impact on patterning precision only at the very weak activation level. It is enough for mutually inhibitory genes C and D to interpret the general patterning blueprint from genes A and B, once the activation degree $\lambda_{5,6}$ is beyond 2×$10^{-7}$. Fig. 7g suggests that the slow or even no diffusion of II system is required for the highly precise patterning.

Fig. 7 also shows that a RD system can become insensitive to external perturbations and generate robust patterns if boundary conditions such as the fixed boundary conditions are set to be non-homogenous with respect to the kinetic steady state [51]. The fixed boundary conditions needed to generate reproducible patterns may correspond to the highly encoded cell fates in animal development [52].

In a recent *in-vivo* study, the combination of AI and II mechanisms consisting of the Bmp-Sox9-Wnt gene network was observed in the digit patterning of mouse (Fig. 8) [53]. Wnt was pointed out to be essential to repress Sox9 in the interdigital regions, which actually suggests the necessity of the II mechanism consisting of Wnt and Sox9 in precise patterning of this process (Fig. 8c). Note that in the current Bmp-Sox9-Wnt gene network, activator Bmp is required to diffuse faster than inhibitor Wnt to generate periodic patterns (Fig. 8d). Nevertheless, this multigene network still follows the classical long-range inhibition in Turing models, since activator Bmp actually provides an inhibitory role on Wnt whereas inhibitor Wnt plays the promoting role on Bmp. Thereby, as analyzed earlier, we can conclude that elevating the difference in self-suppression rates of Wnt and Bmp ($\gamma_2 - \gamma_1$) and decreasing the coupling degree between them (i.e. $|\lambda_1\lambda_2|$ here) will also contribute to the repression of Sox9 (i.e. the flattening distribution). In addition, the current genetic topology network is hard to produce multi-stable patterns like zebra stripes that are established with the gene network in Fig. 7.

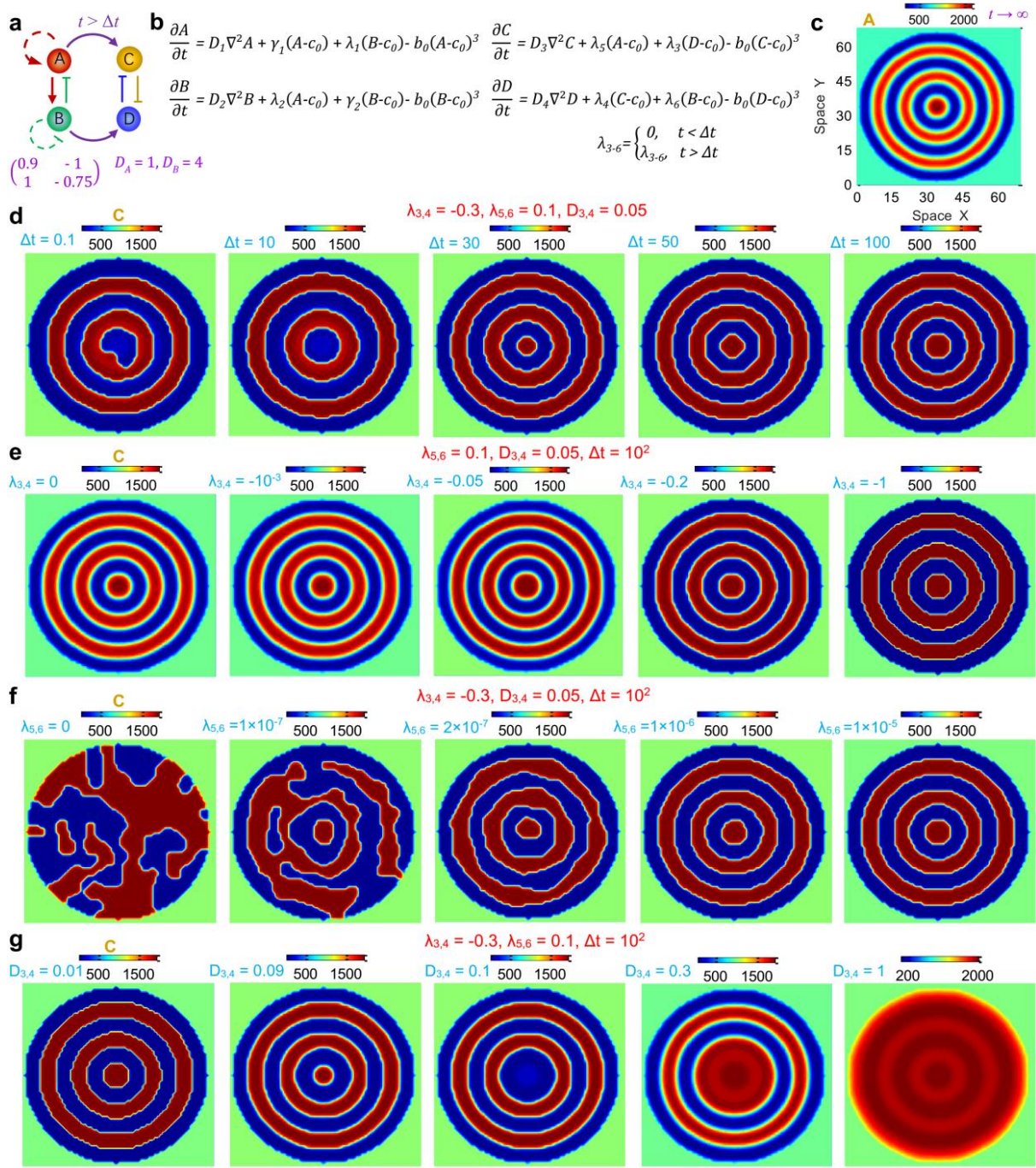

**Figure 7. Bi-stable stripe formation of the sequential AI-II mechanism on stationary 2D domains.** (**a**) Genetic interaction network of the AI-II system. (**b**) Mathematical formulation of the gene network. (**c**) The final spatial distribution of the chemical A. (**d**)-(**g**) The terminal distribution of the chemical C. Fixed boundary conditions are used. $c_0 = 10^3$ is used in all cases.

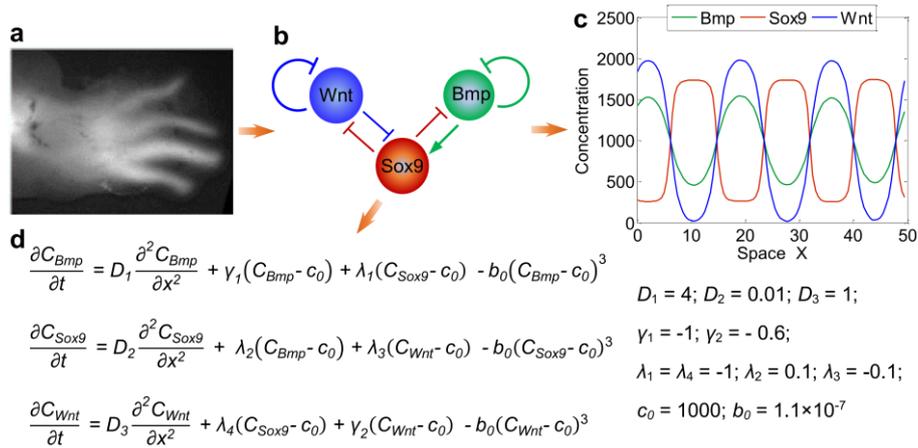

**Figure 8. Digit patterning mechanism identified in mouse.** (**a**) Experimental image [53]. (**b**) Bmp-Sox9-Wnt gene interaction network. (**c**) Our 1D simulation of the gene network (periodic boundary conditions are used). (**d**) Mathematical formulation of the genetic interaction network.

Fig. 9 shows pattern formation of the semi-inhibitor mechanism with equal self-proliferation rates of chemicals. It is observed again that higher self-proliferation rates are able to enhance the patterning precision. Interestingly, with the increment in the self-proliferation rate, the wave length is observed to increase. However, this result violates the theory derived as the Eq. (19), which shows an independence of wave length on the self-proliferation rate when the self-proliferation rates of the two chemicals are equal. This situation suggests that on the 2D space, the patterning principles obtained from 1D system may be altered. The reason is possibly the additional instability originating from a higher dimensional space.

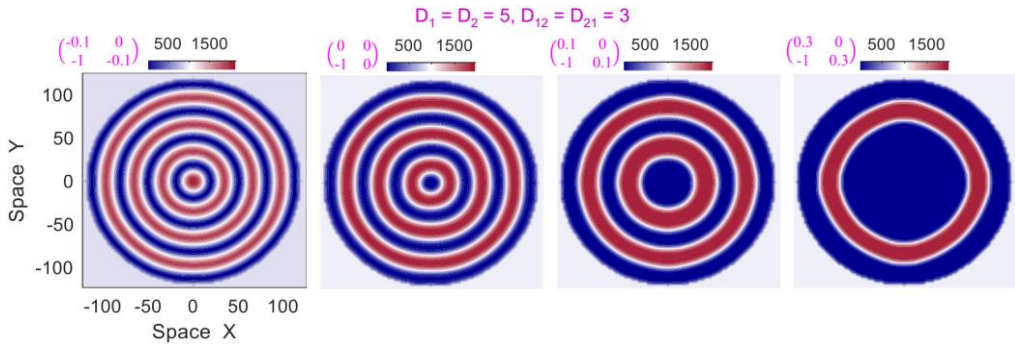

**Figure 9. Precise pattern formation of the semi-activator mechanism in the presence of cross diffusion on stationary 2D domains.** Patterns are produced with the 2D form of Eq. (14) that excludes the effect of domain growth. Fixed boundary conditions are used.

## 4. Summary

Biological pattern formation is widely believed to be the outcome of complex interactions among reacting and diffusing chemicals within biological systems. Local self-activation and lateral long range inhibition has been derived from reaction-diffusion models (Turing models) and achieved success in developmental biology, since it can explain the self-regulation and scaling of biological patterning. In this article, we show that the single AI mechanism long-range inhibition or long-range activation, the semi-inhibitor mechanism and the II mechanism are all able to self-organize into patterns. When cross diffusion is involved, there will be no requirement of the self-enhancement and the different diffusion of chemicals. We also show that the self-organization mechanism is generally unable to be extended to the single AA mechanism. Another obvious problem of Turing models is the low precision, which have made many researchers reluctant to believe the RD mechanism. Here, we propose that the problem does not lie in the diffusion itself but in the single selection of AI mechanism with improper topology parameters. The proper tuning of genetic interaction parameters and the sequential combination among AI, II and AA mechanisms are the likely solutions. Particularly, adding an II system after an AI one is observed to be able to generate patterns with bi-stability. Moreover, we also observe that the patterning principles that are applicable to a low dimensional space may not hold for a high dimensional space, which possibly results from the instability due to the higher degree of spatial freedom of the high dimensional space. The RD mechanism can be involved in the mechanism of precise developmental patterning. However, the direct involvement of II and AA mechanisms in biological patterning, which is based on diffusing activators and inhibitors, largely remains to be disclosed.

The Turing mechanism has been widely believed to be the main mechanism of biological pattern formation over many levels, from intracellular patterning (like cell division [1,54,55] and the gradient formation of intracellular proteins [56]) to animal development (like the digit formation [53] and the somitogenesis [5]) and to ecological evolutions [57]. With the further development of biology, we have reason to believe that more and more evidences will support the RD mechanism. However, we highlight that the RD mechanism is not the only one which can produce periodic patterns. Since the tissue-level patterning in organisms often emerges from the events which happed inside single cells [1], to clearly understand how the subcellular events influence the behaviors at the tissue level, advanced multiscale models with the links from molecular to tissue levels are required [43].

## References

1. Kretschmer, S. & and Schwille, P. Pattern formation on membranes and its role in bacterial cell division. *Curr. Opin. Cell Biol.* **38**, 52-59 (2016).
2. Li, R. & Bowerman, B. Symmetry Breaking in Biology. *Cold Spring Harb Perspect Biol.* **2**, a003475 (2010).
3. Turing, A. M. The chemical basis of morphogenesis. *Philos. Trans. R. Soc. Lond. B Biol. Sci.* **237**, 37-72 (1952).
4. Maini, P. K., Baker, R. E. & Chuong, C. The Turing model comes of molecular age. *Science* **314**, 1397-1398 (2006).
5. Lewis, J. From signals to patterns: space, time, and mathematics in developmental biology. *Science* **322**, 399-403 (2008).
6. Hazen, R. M. The emergence of patterning in life's origin and evolution. *Int. J. Dev. Biol.* **53**, 683-692 (2009).
7. Schweisguth, F. & Corson, F. Self-organization in pattern formation. *Dev. Cell* **49**, 659-677 (2019).
8. Rogers, K. W. & Schier, A. F. Morphogen gradients: from generation to interpretation. *Annu. Rev. Cell Dev. Biol.* **27**, 377-407 (2011).
9. McGuigan, A. P. & Javaherian, S. Tissue patterning: translating design principles from in vivo to in vitro. *Annu. Rev. Biomed. Eng.* **18**, 1-24 (2016).
10. Koch, A. J. & Meinhardt, H. Biological pattern formation: from basic mechanisms to complex structures. *Rev. Mod. Phys.* **66**, 1481-1507 (1994).
11. Roth, S. Mathematics and biology: a Kantian view on the history of pattern formation theory. *Dev. Genes Evol.* **221**, 255-279 (2011).
12. Bolouri, H. Embryonic pattern formation without morphogens. *Bioessays* **30**, 412-417 (2008).
13. Maini, P. K. et al. Turing's model for biological pattern formation and the robustness problem. *Interface Focus* **2**, 487-496 (2012).
14. Ingalls, B. P. Mathematical modeling in systems biology: an introduction. MIT Press: Cambridge, MA, USA, 2012; p. 386.

15. Keller, E. F. & Segel, L. A. Initiation of slime mold aggregation viewed as an instability. *J. Theor. Biol.* **26**, 399-415 (1970).
16. Payne, S. et al. Temporal control of self-organized pattern formation without morphogen gradients in bacteria. *Mol. Syst. Biol.* **9**, 697 (2013).
17. Lee, S. S. & Gaffney, E. A. Aberrant behaviours of reaction diffusion self-organisation models on growing domains in the presence of gene expression time delays. *Bull. Math. Biol.* **72**, 2161-2179 (2010).
18. Gierer, A. & Meinhardt, H. A theory of biological pattern formation. *Kybernetik* **12**, 30-39 (1972).
19. Gaffney, E. A. & Monk, N. A. M. Gene expression time delays and Turing pattern formation systems. *Bull. Math. Biol.* **68**, 99-130 (2006).
20. Alon, U. Network motifs: theory and experimental approaches. *Nat. Rev. Genet.* **8**, 450-461 (2007).
21. Lee, S. S., Gaffney, E. A. & Monk, N. A. M. The influence of gene expression time delays on Gierer-Meinhardt Pattern formation systems. *Bull. Math. Biol.* **72**, 2139-2160 (2010).
22. Wang, X. & Harrison, A. A general principle for spontaneous genetic symmetry breaking and pattern formation within cell populations. *J. Theor. Biol.* **526**, 110809 (2021).
23. Wang, X. & Bai, D. Self-organization principles of cell cycles and gene expressions in the development of cell populations. *Adv. Theor. Simul.* **4**, 2100005 (2021).
24. Segel, L. A. & Jackson, J. L. Dissipative structure: An explanation and an ecological example. *J. Theor. Biol.* **37**, 545-559 (1972).
25. Murray, J. D. Mathematical Biology. Second Ed.. Interdisciplinary Applied Mathematics. New York: Springer-Verlag, 1993.
26. Edelstein-Keshet, Leah. Mathematical Models in Biology. Classics in Applied Mathematics. Society for Industrial and Applied Mathematics, 2005. https://doi.org/10.1137/1.9780898719147.
27. Smith, S. and Dalchau, N. Beyond activator-inhibitor networks: the generalised Turing mechanism. arXiv:1803.07886v1 (2018)
28. Marcon, L. et al. High-throughput mathematical analysis identifies Turing networks for patterning with equally diffusing signals. eLife **5**, e14022 (2016).
29. Langhe, S. P. D et al. Dickkopf-1 (DKK1) reveals that fibronectin is a major target of Wnt signaling in branching morphogenesis of the mouse embryonic lung. *Dev. Biol.* **277**, 316-331 (2005).
30. Augustin, R. et al. Dickkopf related genes are components of the positional value gradient in Hydra. *Dev. Biol.* **296**, 62-70 (2006).
31. Sick, S. et al. WNT and DKK determine hair follicle spacing through a reaction-diffusion mechanism. *Science* **314**, 1447-1450 (2006).
32. Thisse, B., Wright, C. V. & Thisse, C. Activin- and Nodal-related factors control antero–posterior patterning of the zebrafish embryo. *Nature* **403**, 425-428 (2000).
33. Nakamura, T. et al. Generation of robust left-right asymmetry in the mouse embryo requires a self-enhancement and lateral-inhibition system. *Dev. Cell* **11**, 495-504 (2006).
34. Plikus, M. V. et al. Cyclic dermal BMP signalling regulates stem cell activation during hair regeneration. *Nature* 451, 340-344 (2008).
35. Kondo, S. & Miura, T. Reaction-diffusion model as a framework for understanding biological pattern formation. *Science* **329**, 1616-1620 (2010).
36. Kondo, S. & Asai, R. A reaction-diffusion wave on the skin of the marine angelfish Pomacanthus. *Nature* **376**, 765-768 (1995).
37. Crampin, E. J., Gaffney, E. A. & Maini, P. K. Reaction and diffusion on growing domains: scenarios for robust pattern formation. *Bull. Math. Biol.* **61**, 1093-1120 (1999).
38. Crampin, E. J., Hackborn, W. W. & Maini, P. K. Pattern Formation in Reaction–Diffusion Models with Nonuniform Domain Growth. *Bull. Math. Biol.* **64**, 747-769 (2002).
39. Vanag, V. K. & Epstein, I. R. Cross-diffusion and pattern formation in reaction-diffusion systems. *Phys. Chem. Chem. Phys.* **11**, 897-912 (2009).
40. Madzvamuse, A., Gaffney, E. A. & Maini, P. K. Stability analysis of non-autonomous reaction-diffusion systems: the effects

of growing domains. *J. Math. Biol.* **61**, 133-164 (2010).

41. Madzvamuse, A. & Barreira, R. Exhibiting cross-diffusion-induced patterns for reaction-diffusion systems on evolving domains and surfaces. *Phys. Rev. E* **90**, 043307 (2014).

42. Madzvamuse, A., Ndakwo, H. S. & Barreira, R. Cross-diffusion-driven instability for reaction-diffusion systems: analysis and simulations. *J. Math. Biol.* **70**, 709-743 (2015).

43. Kicheva, A., Cohen, M. & Briscoe, J. Developmental pattern formation: insights from physics and biology. *Science* **338**, 210-212 (2012).

44. Wolpert, L. Positional information and the spatial pattern of cellular differentiation. *J. Theor. Biol.* **25**, 1-47 (1969).

45. Green, J, B., New, H. V. & Smith, J. C. Responses of embryonic Xenopus cells to activin and FGF are separated by multiple dose thresholds and correspond to distinct axes of the mesoderm. *Cell* **71**, 731-739 (1992).

46. Ericson, J. et al. Pax6 controls progenitor cell identity and neuronal fate in response to graded Shh signaling. *Cell* **90**, 169-180 (1997).

47. Stathopoulos, A. & Levine, M. Dorsal gradient networks in the Drosophila embryo. *Dev. Biol.* **246**, 57-67 (2002).

48. Tabata, T. & Takei, Y. Morphogens, their identification and regulation. *Development* **131**, 703-712 (2004).

49. Ashe, H. L. BMP signalling: synergy and feedback create a step gradient. *Curr. Biol.* **15**, R375-R377 (2005).

50. Arcuri, P. & Murray, J. D. Pattern sensitivity to boundary and initial conditions in reaction-diffusion models. *J. Math. Biol.* **24**, 141-165 (1986).

51. Barrass, I., Crampin, E. J. & Maini, P. K. Mode transitions in a model reaction-diffusion system driven by domain growth and noise. *Bull. Math. Biol.* **68**, 981-995 (2006).

52. Smith, J., Theodoris, C. & Davidson, E. H. A gene regulatory network subcircuit drives a dynamic pattern of gene expression. *Science* **318**, 794-797 (2007).

53. Raspopovic, J. et al. Digit patterning is controlled by a Bmp-Sox9-Wnt Turing network modulated by morphogen gradients. *Science* **345**, 566-570 (2014).

54. Akiyama, M., Tero, A. & Kobayashi, R. A mathematical model of cleavage. *J. Theor. Biol.* **264**, 84-94 (2010).

55. Kretschmer, S. & Schwille, P. Pattern formation on membranes and its role in bacterial cell division. *Curr. Opin. Cell Biol.* **38**, 52-59 (2016).

56. Halatek, J., Brauns, F. & Frey, E. Self-organization principles of intracellular pattern formation. *Philos. Trans. R. Soc. Lond. B Biol. Sci.* **37**, 201701073 (2018).

57. Rietkerk, M. & van de Koppel, J. Regular pattern formation in real ecosystems. *Trends Ecol. Evol.* **23**, 169-175 (2008).

**Acknowledgements:** This work is supported by Zhejiang University, National Natural Science Foundation of China and the research builds on research on the calibration of gene expression experiments funded by the UK BBSRC (BB/E001742/1). The authors are grateful to comments and improvement suggestions from reviewers.

**Author contributions**: X.W. conceived this research, developed models, implemented theoretical calculations and data analysis, and wrote the paper. A.H. evaluated the article, provided revision suggestions and contributed to the writing. All authors participated in discussion.

**Conflict of interest**: The authors declare no conflict of interest.